\documentstyle[twoside,fleqn,npb,epsfig]{article}
\newcommand{\AmS}{{\protect\the\textfont2
  A\kern-.1667em\lower.5ex\hbox{M}\kern-.125emS}}

\def    \pl     #1#2#3{{\it Phys. Lett.} {\bf #1}(19#2)#3}

\def    \pr     #1#2#3{{\it Phys. Rev.} {\bf #1}(19#2)#3}

\def    \eup    #1#2#3{{\it Eur. Phys. J.} {\bf #1}(19#2)#3}
\def    \hepph  #1 {{\tt hep-ph/#1}}
\def    \hepex  #1 {{\tt hep-ex/#1}}
\def\beq{\begin{equation}}
\def\eeq{\end{equation}}
\def\beqn{\begin{eqnarray}}
\def\eeqn{\end{eqnarray}}
\def\sss{\scriptscriptstyle}
\def\pt{p_{\sss T}}

\def\ptg{p_{{\sss T}\gamma}}
\def\ptj{p_{{\sss T}j}}

\def\ep{\epsilon}

\begin{document}

\twocolumn[
\vskip -3cm
~

\flushright{
        \begin{minipage}{2.9cm}
        CERN-TH/99-158\hfill \\
        hep-ph/9905546\hfill \\
        \end{minipage}        }

% declarations for front matter
\vskip 5.0em
\flushleft{\Large Isolated-photon production in polarized 
hadronic collisions}~\footnotemark
\vskip\baselineskip
\flushleft{Stefano Frixione$^{\rm a}$}
\vskip\baselineskip
\flushleft{$^{\rm a}$CERN, TH division, Geneva, Switzerland}

\begin{center}
\begin{minipage}{160mm}
                \parindent=10pt
{\small 
After a short discussion on the definition of isolated-photon cross
sections in perturbative QCD, I present phenomenological predictions
relevant for polarized hadron-hadron collisions in the RHIC energy
range. The possibility of measuring $\Delta g$ is investigated.
}
                \par
                \end{minipage}
                \vskip 2pc \par
\end{center}
]
\footnotetext{Talk given at DIS99, 19-23 April 1999, Zeuthen, DE.}

% typeset front matter (including abstract)
%\thispagestyle{empty}

\section{Photon isolation in perturbative QCD}

The production of photons in hadronic collisions is quite interesting
under two different aspects. Since the photon couples directly only
to quarks, photon signals can be used to study the hard dynamics
in a cleaner way with respect to processes where only hadrons are 
involved; the obvious drawback is that prompt-photon cross sections
are much smaller than - say - jet cross sections.
Secondly, photon data constitute an unique tool for pinning down
the gluon density at intermediate and large $x$, since the number 
of partonic processes involved is smaller with respect to other 
processes which are equally or more sensitive to the gluon density. 
Unfortunately, the real situation is much worse than that described above.
The main reason is that in a complicated environment, like that
arising in high-energy hadronic collisions, there are lots of photons
around, mainly coming from the decay $\pi^0\!\to\!\gamma\gamma$.
This is a huge background, since the two photons are often
detected as a single one. However, the signature of photons coming
from hadron decays is rather different from that of prompt photons,
the mother hadron being usually surrounded by other hadrons. Therefore,
photons originating from decays have the same direction of a relatively 
large number of hadrons. It follows that an efficient way of selecting
prompt-photon events is that of requiring the photon to be {\it isolated}
from hadron tracks in the detector.
There is also a further complication: photons can originate from 
fluctuations of hadrons with the same quantum numbers. Perturbative
QCD is not able to deal with hadrons, and the fluctuation of a hadron
into a photon is effectively described by the (non-perturbative)
fragmentation of a parton into a photon. The cross section for
the process $H_1 H_2\!\to\!\gamma\!+\!X$ is therefore written as follows
\beqn
&&\!\!\!\!\!\!\sigma_\gamma^{(H_1 H_2)}=
f_i^{(H_1)}\otimes f_j^{(H_2)}\otimes\hat{\sigma}_{ij;\gamma}
\nonumber \\&&\phantom{\!\!\!\!\!\!\sigma_{H_1 H_2;\gamma}}
+f_i^{(H_1)}\otimes f_j^{(H_2)}\otimes\hat{\sigma}_{ij;k}
\otimes D_\gamma^{(k)}.
\label{factth}
\eeqn
The first term in the RHS of this equation describes the production
of photons in the hard process $ij\!\to\!\gamma\!+\!X$ (direct mechanism),
while in the second term (fragmentation mechanism) the hard process
is $ij\!\to\!k\!+\!X$, with the parton $k$ eventually fragmenting into 
a photon, in a way parametrized by the fragmentation function (FF)
$D_\gamma^{(k)}$, which is not calculable in perturbation theory but 
is universal. Of course, the fragmentation mechanism does not give any clean 
information on the parton dynamics. However, the photon produced in
this way will be close to the hadron remnants; therefore, the 
isolation condition will reduce the contribution of the 
fragmentation mechanism to the physical cross section.

The key question is: how much? In fact, in perturbative QCD the direct
and fragmentation parts are very closely related, since there are 
divergences that cancel only when these two contributions are
summed together. From the phenomenological point of view this is
annoying, since it is difficult to estimate the impact of the
very poorly known FFs on prompt-photon analyses.
However, there exists at least one isolation prescription~\cite{iso98} 
which is such that the contribution of the fragmentation part to the 
physical observables is exactly zero, and still the cross section is 
finite at all orders in perturbation theory. The isolation condition is
as follows: drawing a cone of half-angle $R_0$ around the photon 
axis (in the $\eta -\phi$ plane), and denoting by $E_{T,had}(R)$ 
the total amount of transverse hadronic energy inside a cone of 
half-angle $R$, the photon is isolated if the following inequality 
is satisfied:
\beq
E_{T,had}(R)\le \ep_\gamma p_{{\sss T}\gamma} {\cal Y}(R),
\label{iscond}
\eeq
for all $R\le R_0$. Here, $p_{{\sss T}\gamma}$ is the transverse momentum of 
the photon. The function ${\cal Y}$ can be rather freely chosen, provided
that it vanishes fast enough for $R\to 0$. A sensible choice is the
following:
\beq
{\cal Y}(R)=\left(\frac{1-\cos R}{1-\cos R_0}\right)^n,
\;\;\;\;n=1,\;\;\;\;\ep_\gamma=1.
\label{isfun}
\eeq
Notice that the ordinary cone isolation can be recovered from
eqs.~(\ref{iscond}) and~(\ref{isfun}) by setting $n=0$ and
$\ep_\gamma=\ep_c$. I also stress that the same prescription can be
applied, with trivial modifications, to any other type of hard
collisions. It has been shown that this definition of isolated photon
leads to an infrared-safe cross section in perturbative QCD. More
details can be found in ref.~\cite{iso98}.

\section{Isolated photons at RHIC}

\begin{figure}[htb]
\vspace{9pt}
\epsfig{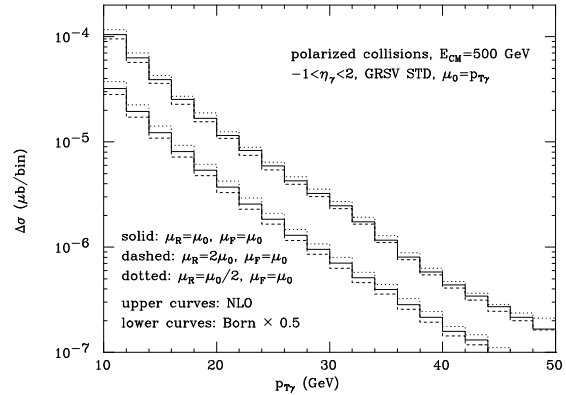}
\caption{Scale dependence of the $\pt$ spectrum.}
\label{fig:scale}
\end{figure}

The arguments of the previous section apply to both polarized and
unpolarized collisions. In the following, I will present
phenomenological predictions relevant for isolated-photon production
in polarized hadronic collisions, in the RHIC energy range
($\sqrt{S}=200\div 500$ GeV). The results are accurate to NLO in QCD.
Studies of unpolarized hadronic collisions, for higher center-of-mass
energies, have been presented elsewhere~\cite{Vancouver}. The
computer code used here is a modification of that of
ref.~\cite{Vancouver}. I start from considering the perturbative
stability of isolated-photon cross sections. This is an important
issue to study, since it is well known that the isolation cuts result
in an imperfect cancellation of soft singularities, which might lead
to unreliable perturbative results. As customary when a NNLO
calculation is lacking, the perturbative stability is studied looking
at the variation of the cross section induced by the variation of the
factorization and renormalization scales with respect to a default
scale $\mu_0$. A sample result is given in fig.~\ref{fig:scale}, where
the $\pt$ spectrum of the photon is presented. From the figure, we can
see that the size of the radiative corrections is moderate, and the
scale dependence is milder at NLO than at LO. Only the renormalization
scale has been varied; the dependence of the cross section upon the
factorization scale is small at LO, and is basically zero at NLO.
%%%%%%%%%%%%%%%%%%%%%%%%%%%%%%%%%%%%%%%%%%%%%%%%%%%%%%%%%%%%%%%%%%%%%%
%%%%%%%%%%%%%%%%%%%%%%%%%%%%%%%%%%%%%%%%%%%%%%%%%%%%%%%%%%%%%%%%%%%%%%
Other observables have been studied as well, and they display the same
pattern of scale dependence as the $\pt$ spectrum presented in
fig.~\ref{fig:scale}. Results for unpolarized scattering (of relevance
since the measurable quantities are asymmetries rather than absolute
cross sections) have also been obtained, and they are consistent with
their polarized counterpart. In summary, it seems that the
perturbation series in the case of isolated photons defined as in
eq.~(\ref{iscond}) is under control, and NLO results can be used to
perform a sensible phenomenological study.

\begin{figure}[htb]
\vspace{9pt}
\epsfig{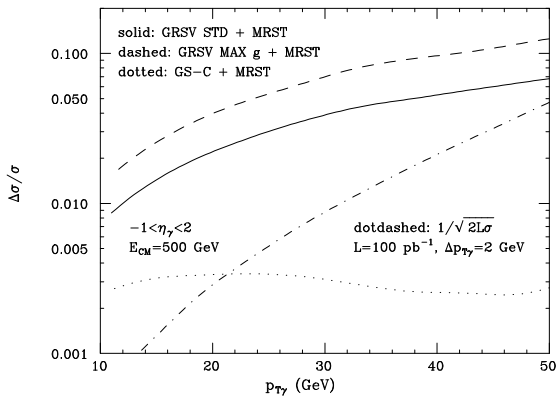}
\caption{Asymmetry versus $\ptg$. The minimum observable
asymmetry is also shown.}
\label{fig:asypt}
\end{figure}

The main goal of collecting isolated-photon data at RHIC will that
of measuring the gluon density $\Delta g$ in the polarized proton.
This quantity is to a large extent unknown at present, and the
various available parametrizations are very different from each 
other. The isolated-photon cross section is much smaller than,
for example, the single-inclusive jet cross section (using GRSV STD 
densities~\cite{GRSV}, default scales and $\sqrt{S}=500$~GeV, 
the former is $2.9~10^{-4}~\mu b$ 
in the range $\ptg\!>\!10$~GeV, $-1\!<\!\eta_\gamma\!<\!2$, while the latter 
is $3.4~10^{-1}~\mu b$ in the range $\ptj\!>\!10$~GeV, $-1\!<\!\eta_j\!<\!1$).
Therefore, although in principle a measurement of $\Delta g$ performed
with photon events is cleaner and simpler (the cross section at LO
depends linearly upon $\Delta g$, while in the jet case the dependence
is quadratic), it remains to see whether in practice the sensitivity
to $\Delta g$ will be large enough to allow a discrimination between
the various parametrizations. This issue is studied in fig.~\ref{fig:asypt},
%%%%%%%%%%%%%%%%%%%%%%%%%%%%%%%%%%%%%%%%%%%%%%%%%%%%%%%%%%%%%%%%%%%%%%
%%%%%%%%%%%%%%%%%%%%%%%%%%%%%%%%%%%%%%%%%%%%%%%%%%%%%%%%%%%%%%%%%%%%%%
where the $\ptg$ dependence of the longitudinal asymmetry is presented.
The asymmetry has been calculated for three different sets of polarized
parton densities: GRSV STD, GRSV MAXg~\cite{GRSV}, and GS-C~\cite{GSC}. 
The latter
two sets represent quite an extreme choice: all the other available
densities return an asymmetry whose value is larger than that obtained
with GS-C and smaller than that obtained with GRSV MAXg. For the three
asymmetries, the MRST set~\cite{MRST} has been adopted to compute the
unpolarized cross section. Fig.~\ref{fig:asypt} also displays the
minimum observable value of the asymmetry:
\beq
\left({\cal A}_{\pt}\right)_{min}=
\frac{1}{P^2} \frac{1}{\sqrt{2\sigma {\cal L}\ep}} ,
\label{Amin}
\eeq
where ${\cal L}$ is the integrated luminosity, $P$ is the polarization
of the beam, and the factor $\ep\le 1$ accounts for experimental
efficiencies; $\sigma$ is the unpolarized cross section integrated
over a small range in transverse momentum ($\pt$ bin). The quantity
defined in eq.~(\ref{Amin}) is plotted (dot-dashed line) in
fig.~\ref{fig:asypt}, for $\ep=1$, $P=1$, ${\cal L}=100$~pb$^{-1}$ and
a $\pt$-bin size of 2 GeV. Taking into account that RHIC expects to
collect a luminosity much higher than 100~pb$^{-1}$, that a value of
$P=0.7$ is realistic, and hoping for detector performances such that
$\ep\simeq 1$, we see from the figure that the asymmetry is measurable
at the lowest $\ptg$ values, regardless of the polarized density set
adopted. The measurement is increasingly difficult for larger $\ptg$,
probably becoming impossible for transverse momenta of the order of
$40\div 50$~GeV. At $\sqrt{S}=200$ GeV asymmetries are larger, but
$({\cal A}_{\pt})_{min}$ is also larger, and the conclusions are
unchanged.

In summary, I presented NLO results for photon production in
polarized hadronic collisions, the photon being isolated in a way such
that the cross section does not depend upon the fragmentation part. A
more complete discussion will be given elsewhere~\cite{FV}.
Phenomenological predictions have been given for the RHIC energy
range. It has been shown that the perturbative series is under
control, and that isolated photon production at RHIC will be a viable
tool in the study of the polarized parton densities.

\end{document}